\documentclass[%
reprint,
 amsmath,amssymb,
 aps,
 twocolumn,
pra,
]{revtex4-1}

\usepackage[T1]{fontenc}
\usepackage[dvipdfmx]{graphicx}
\usepackage{dcolumn}
\usepackage{bm}
\usepackage[
colorlinks=true,
linkcolor=blue,
filecolor=blue,
citecolor=blue,
urlcolor=blue
]{hyperref}

\usepackage{comment}

\usepackage{color}
\usepackage{braket}
\usepackage[caption=false]{subfig}
\usepackage{amsmath}
\usepackage{mathtools}

\usepackage{lineno}

\begin{document}

\title{Continuous-variable quantum kernel method\\ on a programmable photonic quantum processor}

\author{Keitaro Anai}%
\affiliation{%
 Department of Applied Physics, School of Engineering, The University of Tokyo,\\
 7-3-1 Hongo, Bunkyo-ku, Tokyo 113-8656, Japan
}%
\author{Shion Ikehara}
\affiliation{%
 Department of Applied Physics, School of Engineering, The University of Tokyo,\\
 7-3-1 Hongo, Bunkyo-ku, Tokyo 113-8656, Japan
}%
\author{Yoshichika Yano}
\affiliation{%
 Department of Applied Physics, School of Engineering, The University of Tokyo,\\
 7-3-1 Hongo, Bunkyo-ku, Tokyo 113-8656, Japan
}%
\author{Daichi Okuno}
\affiliation{%
 Department of Applied Physics, School of Engineering, The University of Tokyo,\\
 7-3-1 Hongo, Bunkyo-ku, Tokyo 113-8656, Japan
}%
\author{Shuntaro Takeda}
\email{takeda@ap.t.u-tokyo.ac.jp}
\affiliation{%
 Department of Applied Physics, School of Engineering, The University of Tokyo,\\
 7-3-1 Hongo, Bunkyo-ku, Tokyo 113-8656, Japan
}%





\date{\today}
\begin{abstract}
\noindent 
Among various quantum machine learning (QML) algorithms, the quantum kernel method has especially attracted attention due to its compatibility with noisy intermediate-scale quantum devices and its potential to achieve quantum advantage. This method performs classification and regression by nonlinearly mapping data into quantum states in a higher dimensional Hilbert space. Thus far, the quantum kernel method has been implemented only on qubit-based systems, but continuous-variable (CV) systems can potentially offer superior computational power by utilizing its infinite-dimensional Hilbert space. Here, we demonstrate the implementation of the classification task with the CV quantum kernel method on a programmable photonic quantum processor. We experimentally prove that the CV quantum kernel method successfully classifies several datasets robustly even under the experimental imperfections, with high accuracies comparable to the classical kernel. This demonstration sheds light on the utility of CV quantum systems for QML and should stimulate further study in other CV QML algorithms.
\end{abstract}

\maketitle


\section{Introduction}
Recently, studies have been intense to realize large-scale fault-tolerant quantum computing, but at present, we can only access noisy intermediate-scale quantum (NISQ) devices~\cite{ths:NISQ_review}. Even with such devices, the realization of quantum advantage for impractical problems has already been reported~\cite{ths:Q_adv1, ths:Q_adv2, ths:Q_adv3}. To go beyond this, there has been much interest in achieving quantum advantage for practical problems. One of the most promising candidates for this goal is quantum machine learning (QML)~\cite{ths:QML_review, ths:QKM_review}. Thus far, a wide variety of QML for NISQ devices has been proposed theoretically for qubit-based systems~\cite{ths:QML_thy_ad4, ths:QML_thy_ad1, ths:QML_thy_ad2, ths:QML_thy_ad3} and demonstrated experimentally on several physical platforms~\cite{ths:QML_exp_ad1, ths:DV_QKM_sycamore, ths:DV_QML_handai, ths:QKM_exp_ad1, ths:QKM_exp_ad2}.\par

Among such QML algorithms, a quantum kernel method has especially attracted attention due to its simplicity of implementation~\cite{ths:DV_QML_handai, ths:DV_QKM_sycamore, ths:QKM_exp_ad1, ths:QKM_exp_ad2} and its potential to achieve quantum advantage. In general, a classical kernel method aims to perform tasks such as classification and regression by nonlinearly mapping data into a higher-dimensional Hilbert space~\cite{ths:RKHS}. On the other hand, the quantum kernel method utilizes quantum states as a nonlinear mapping~\cite{ths:CV_QML_feature}. The quantum kernel method is expected to be advantageous compared with the classical one because the mapping to complex quantum states potentially has the ability to recognize classically intractable complex patterns. Refs.~\cite{ths:QML_adv_1, ths:QML_adv2, ths:GeneralAdvQML} have proved that this approach has rigorous advantage over classical computation in specific tasks.

In contrast to the qubit systems, there have been few theoretical proposals~\cite{ths:CV_QML_feature, ths:CV_QKM_nonClass, ths:CV_QML_OurProp} and no experimental implementations of the continuous-variable (CV) quantum kernel method, although CV quantum computing can potentially offer superior computational power in the NISQ architecture. CV quantum systems can handle infinite-dimensional Hilbert space with only one mode in contrast to qubit systems with two-dimensional Hilbert space per one qubit, which indicates that we can map data into a larger Hilbert space with the CV system. It can also easily construct a kernel that has not been studied in the classical kernel method by using quantum states unique to CV systems such as squeezed states~\cite{ths:CV_QML_feature, ths:CV_QML_OurProp}.

In this paper, we successfully implement the classification task with the CV quantum kernel method using a programmable photonic quantum processor. We map given data to phases of squeezed states~\cite{ths:CV_QML_feature, ths:CV_QML_OurProp} via a quantum feature map and experimentally obtain the kernel matrix by measuring the pair-wise inner product of the feature states using a single-mode photonic quantum processor. The obtained kernel is then given to a classical processor that solves a convex quadratic program~\cite{ths:libsvm} and thereby efficiently finds the linear classifier which optimally separates the training data in feature space. By implementing the procedure above, we experimentally prove that the CV quantum kernel method successfully classifies several datasets robustly even under the experimental imperfections, with high accuracies comparable to the classical kernel. Our paper experimentally demonstrates the utility of the CV quantum kernel method. Moreover, our demonstration of CV QML should stimulate the study of the other CV QML methods, such as quantum neural network~\cite{ths:CV_Neural}, quantum reservoir computing~\cite{ths:reservoir}, and other quantum learning algorithms~\cite{ths:CV_linear}.

\begin{figure*}[t]
\includegraphics[width=1\linewidth]{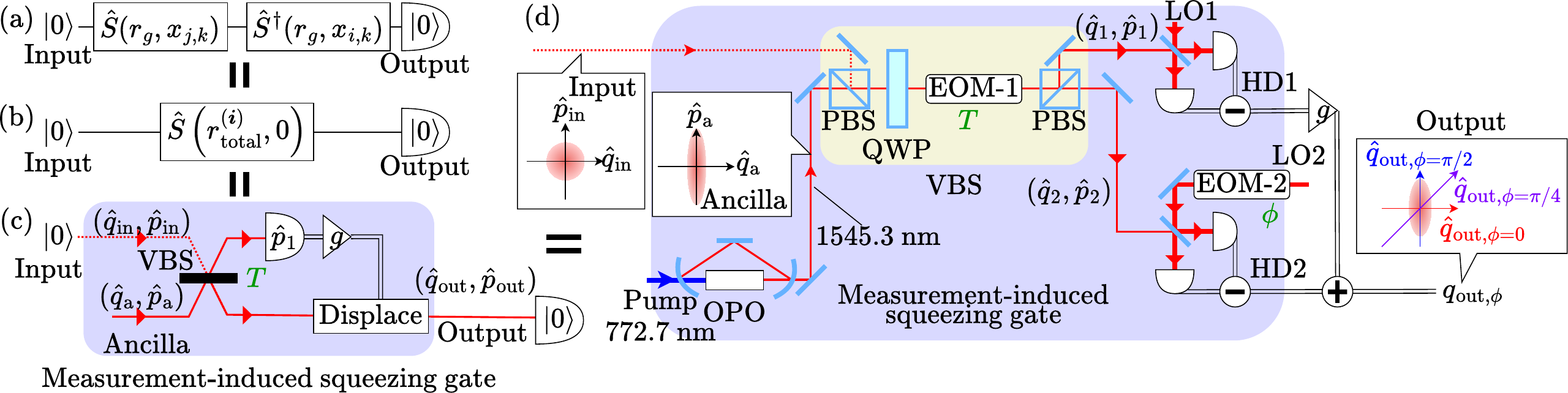}
\centering
\caption{Experimental implementation of the CV quantum kernel method. (a) Conceptual diagram of measurement of our quantum kernel. The input state is a vacuum state. Squeezing operators $\hat{S}^\dagger(r_g,x_{i,k})\hat{S}(r_g,x_{j,k})$ are applied to it for input data $x_{i,k}$ and $x_{j,k}$. Finally, the vacuum component of the output state is measured. (b) Conceptual diagram of our demonstration. Two squeezing operators $\hat{S}(r_g,x_{j,k})$ and $\hat{S}^\dagger(r_g,x_{i,k})$ are combined into one squeezing operator $\hat{S}(r_\mathrm{total}^{(\bm{i})}, 0)$. (c) Optical circuit to implement the circuit shown in Fig.~\ref{fig:setup}(b). ``Displace'' indicates a displacement operation depending on the measurement result of $\hat{p}_1$, $(\hat{q}_i,\hat{p}_i)$ are quadrature amplitudes of mode $i$ whose definitions and expressions are given in the main text, $g$ is a feedforward gain, and $T$ is a transmissivity of the beam splitter. (d) Experimental setup. EOM, electro-optic modulator; HD, homodyne detector; LO, local oscillator; OPO, optical parametric oscillator; PBS, polarizing beam splitter; QWP, quarter-wave plate; VBS, variable beam splitter.}\label{fig:setup}
\end{figure*}

\section{Theory of quantum kernel method}
In this paper, we consider the situation that we are given training datum $\bm{x}_i=(x_{i,1},x_{i,2})\in \mathbb{R}^2 (i=1,\cdots,l)$ and its label $y_i\in\{-1,1\}$. Our goal is to recognize the pattern and predict label $y_\mathrm{test}$ of unseen test datum $\bm{x}_\mathrm{test}$. To achieve this, the classical kernel method encodes $\bm{x}_i$ to a higher-dimensional Hilbert space $\mathcal{H}$ with a feature map $\phi:\mathbb{R}^2\rightarrow\mathcal{H}$. Then, we compute a kernel which measures a similarity of two data and is defined by $K(\bm{x}_i,\bm{x}_j)=\braket{\phi(\bm{x}_i),\phi(\bm{x}_j)}_\mathcal{H}$, where $\braket{\cdot,\cdot}_\mathcal{H}$ is an inner product defined on $\mathcal{H}$. With this kernel, we can construct a classification model and predict the label of unseen data. The classification model is obtained by first solving a convex quadratic program
\begin{equation}\label{eq:dualprob}
\begin{split}
    \min_{\bm{\alpha}}&\quad\frac{1}{2}\bm{\alpha}^TQ\bm{\alpha}-\sum_{i=1}^l\alpha_i\\
    \mathrm{subject\ to}&\quad\bm{y}^T\bm{\alpha}=0,\ 0\leq\alpha_i\leq1,\ i=1,\cdots,l,
\end{split}
\end{equation}
where $Q$ is an $l$ by $l$ positive semidefinite matrix with $Q_{ij}=y_iy_jK(\bm{x}_i,\bm{x}_j)$, $\bm{y}=(y_1,\cdots,y_l)^\mathsf{T}\in\mathbb{R}^l$, and $\bm{\alpha}=(\alpha_1,\cdots,\alpha_l)^\mathsf{T}\in\mathbb{R}^l$. After Eq.~\eqref{eq:dualprob} is solved and $\bm{\alpha}$ is determined, we can predict label $y_\mathrm{test}$ of test datum $\bm{x}_\mathrm{test}$ as
\begin{equation}
    y_\mathrm{pred}=\mathrm{sgn}\left(\sum_{i=1}^ly_i\alpha_iK(\bm{x}_i,\bm{x}_\mathrm{test})+b\right),
\end{equation}
where $\mathrm{sgn}(\cdot)$ is a sign function. Note that $b\in\mathbb{R}$ can be also calculated straightforwardly from $\bm{\alpha}$ and $(\bm{x}_i,y_i)$~\cite{ths:libsvm}.\par

In contrast, the quantum kernel method maps datum $\bm{x}$ to a quantum state to utilize the large-dimensional and quantum Hilbert space. Among several proposals about CV quantum kernels~\cite{ths:CV_QML_feature, ths:CV_QKM_nonClass, ths:CV_QML_OurProp}, we encode data into squeezed states, one of the typical CV quantum states. As for this squeezed-state encoding, two ways of encoding are proposed in Ref.~\cite{ths:CV_QML_OurProp}: amplitude encoding and phase encoding (\textit{squeezing-phase kernel}). Here, we adopt phase encoding because it has a hyperparameter (the squeezing level of the squeezed state) that can be adjusted for given datasets, whereas amplitude encoding does not have such a hyperparameter.\par

A squeezed state can be represented as $\hat{S}(r,\theta)\ket{0}$, where $\hat{S}(r,\theta)=\exp\left((re^{-i\theta}\hat{a}^2-re^{i\theta}(\hat{a}^\dagger)^2)/2\right)$ is the squeezing operator, $\hat{a}$ is an annihilation operator, and $\ket{0}$ is a vacuum state. Real parameters $r$ and $\theta$ determine the squeezing level and the phase, respectively. The squeezing-phase kernel maps datum $\bm{x}$ to squeezing phase $\theta$~\cite{ths:CV_QML_OurProp} and is defined as
\begin{align}
    K_{r_g}(\bm{x}_i,\bm{x}_j)&=\prod_{k=1}^2\kappa_{r_g}(x_{i,k},x_{j,k})\nonumber\\
    &=\prod_{k=1}^2\|\braket{0|\hat{S}^\dagger(r_g,x_{i,k})\hat{S}(r_g,x_{j,k})|0}\|^2.\label{eq:kernel}
\end{align}
\noindent
Here, we call parameter $r_g$ a \textit{gate-squeezing level}. This parameter can be regarded as a hyperparameter which has to be fixed before learning or tuned so that the performance of learning is maximized. We also note that the squeezing-phase kernel is a periodic function with a period of $2\pi$. Such periodic kernels are suitable for the classification of periodic data~\cite{ths:CV_QML_OurProp}. We can also classify non-periodic data by normalizing the data so that they are always within one period of the kernel. This strategy is taken in our experiment and will be described later.

\section{Concept of our implementation} 
To evaluate $\kappa_{r_g}(x_{i,k},x_{j,k})$ in Eq.~\eqref{eq:kernel} with a photonic quantum processor, we first consider the circuit in Fig.~\ref{fig:setup}(a). In this setup, we prepare $\ket{0}$ as an input, implement operation $\hat{S}^\dagger(r_g,x_{i,k})\hat{S}(r_g,x_{j,k})$, and measure the vacuum component of the output state. Operation  $\hat{S}^\dagger(r_g,x_{i,k})\hat{S}(r_g,x_{j,k})$ can be decomposed into $\hat{R}(\phi_1^{(\bm{i})})\hat{S}(r_\mathrm{total}^{(\bm{i})},0)\hat{R}(\phi_2^{(\bm{i})}),\ \bm{i}=(i,j,k)$ with Bloch Messiah decomposition~\cite{ths:bloch}, where $r_\mathrm{total}^{(\bm{i})}$ is determined as
\begin{align}
    e^{2r_\mathrm{total}^{(\bm{i})}}&=\cos^2\Delta^{(\bm{i})}+\cosh(4r_g)\sin^2\Delta^{(\bm{i})}\nonumber\\
        &+\left[\sinh^2(4r_g)\sin^4\Delta^{(\bm{i})}\right.\nonumber\\
        &\left.+4\sinh^2(2r_g)\sin^2\Delta^{(\bm{i})}\cos^2\Delta^{(\bm{i})}\right]^{\frac{1}{2}},\label{eq:r_total}\\
        \Delta^{(\bm{i})}&=\frac{x_{i,k}-x_{j,k}}{2},
\end{align}
\noindent
and $\hat{R}(\phi)=\exp(-i\phi\hat{a}^\dagger\hat{a})$ is a phase-shift operator (see Appendix~\ref{apa} for details). With equation $\hat{R}(\phi)\ket{0}=\ket{0}$, $\kappa_{r_g}(x_{i,k},x_{j,k})$ in Eq.~\eqref{eq:kernel} turns into 
\begin{align}
    \kappa_{r_g}(x_{i,k},x_{j,k})&=\|\braket{0|\hat{R}(\phi_1^{(\bm{i})})\hat{S}(r_\mathrm{total}^{(\bm{i})},0)\hat{R}(\phi_2^{(\bm{i})})|0}\|^2\nonumber\\
    &=\|\braket{0|\hat{S}(r_\mathrm{total}^{(\bm{i})},0)|0}\|^2.\label{eq:KernelExp}
\end{align}
\noindent
Hence, to evaluate $\kappa_{r_g}(x_{i,k},x_{j,k})$ in our experiment, we need only $\hat{S}(r_\mathrm{total}^{(\bm{i})},0)$ instead of $\hat{S}^\dagger(r_g,x_{i,k})\hat{S}(r_g,x_{j,k})$ as shown in Fig.~\ref{fig:setup}(b). Note that from Eq.~\eqref{eq:r_total}, the kernel value is dependent only on the absolute difference of data $|x_{i,k}-x_{j,k}|$, rather than the datum itself.\par

To realize the concept of Fig.~\ref{fig:setup}(b), we next consider the circuit shown in Fig.~\ref{fig:setup}(c). In Fig.~\ref{fig:setup}(c), we use a measurement-induced squeezing gate~\cite{ths:univsq} to perform the squeezing operation $\hat{S}(r_\mathrm{total}^{(\bm{i})},0)$ in a programmable way. The input state is a vacuum state with quadrature amplitudes $(\hat{q}_\mathrm{in},\hat{p}_\mathrm{in})$. The ancillary state is a squeezed state with $\braket{\hat{q}_\mathrm{a}^2}=e^{-2r_\mathrm{a}}$ and $\braket{\hat{p}_\mathrm{a}^2}=e^{2r_\mathrm{a}}$, where $(\hat{q}_\mathrm{a}, \hat{p}_\mathrm{a})$ represents quadrature amplitudes of the ancillary mode. Here, we call parameter $r_\mathrm{a}$ an \textit{ancilla-squeezing level}. These two fields interfere at the beam splitter having variable transmissivity $T$. Then, the homodyne detector measures quadrature amplitude $\hat{p}_1$ of one of the beam-splitter output modes, and measurement outcome $p_1$ is fed forward to the other mode with a certain gain $g$ by displacing the quadrature amplitude by $gp_1$ along the $p$ direction. To let the above-described circuit act as squeezing gate $\hat{S}(r_\mathrm{total}^{(\bm{i})},0)$, the parameters of the optical circuit are determined by $T=\exp(-2r_\mathrm{total}^{(\bm{i})})$ and correspondingly the feedforward gain is set to $g=\sqrt{(1-T)/T}$, in which settings the contribution of the antisqueezed quadrature of ancilla $\hat{p}_\mathrm{a}$ to the gate output is canceled. The quadrature amplitudes of output state $\hat{q}_\mathrm{out,\mathrm{\phi}}=\hat{q}_\mathrm{out}\cos\phi+\hat{p}_\mathrm{out}\sin\phi$ in Fig.~\ref{fig:setup}(c) becomes
\begin{align}
    \hat{q}_{\mathrm{out},\phi}&=e^{-r_\mathrm{total}^{(\bm{i})}}\hat{q}_\mathrm{in}\cos\phi+e^{r_\mathrm{total}^{(\bm{i})}}\hat{p}_\mathrm{in}\sin\phi\nonumber\\
    &-\sqrt{1-\exp(-2r_\mathrm{total}^{(\bm{i})})}\hat{q}_\mathrm{a}\cos\phi,\label{eq:output}
\end{align}
which asymptotically coincides with results of the ideal squeezing gate $\hat{S}(r_\mathrm{total}^{(\bm{i})},0)$ in high squeezing limit $r_{\mathrm{a}}\rightarrow\infty$ in the sense that the variance of the noise term $\sqrt{1-\exp(-2r_\mathrm{total}^{(\bm{i})})}\hat{q}_\mathrm{a}\cos\phi$ approaches zero (details of this derivation are described in Appendix~\ref{apb}).\par

Finally, Fig.~\ref{fig:setup}(d) shows our experimental configuration. This setup realizes the circuit in Fig.~\ref{fig:setup}(c), but some parts of it are modified for experimental convenience. First, the final measurement of the vacuum component of $\hat{S}(r_\mathrm{total}^{(\bm{i})},0)\ket{0}$ is performed by a homodyne detector (HD2) and the analysis of its measurement result. Though a photon detector is straightforward to measure the vacuum component, the homodyne detector allows for the measurement with lower optical loss compared to the photon detector. Second, for experimental simplicity, we perform the feedforward operation not by optical means but by numerical post-processing of HD2's measurement outcome, as shown in Fig.~\ref{fig:setup}(d). This processing yields the same outcome as Eq.~\eqref{eq:output} without any loss of data or shots (see Appendix~\ref{apb} for details; similar tricks can be seen in other works~\cite{ths:OurQAOA, ths:PostProcess2, ths:PostProcess3}). Through the above procedure, we can evaluate $\kappa_{r_g}(x_{i,k},x_{j,k})$ in Eq.~\eqref{eq:kernel}. Hence, the kernel $K_{r_g}(\bm{x}_i,\bm{x}_j)$ can be obtained by using the circuit in Fig.~\ref{fig:setup}(d) twice, evaluating $\kappa_{r_g}(x_{i,k},x_{j,k})$ for $k=1,2$, and then multiplying them.\par

\begin{figure*}[t]
\centering
\includegraphics[width=1\linewidth]{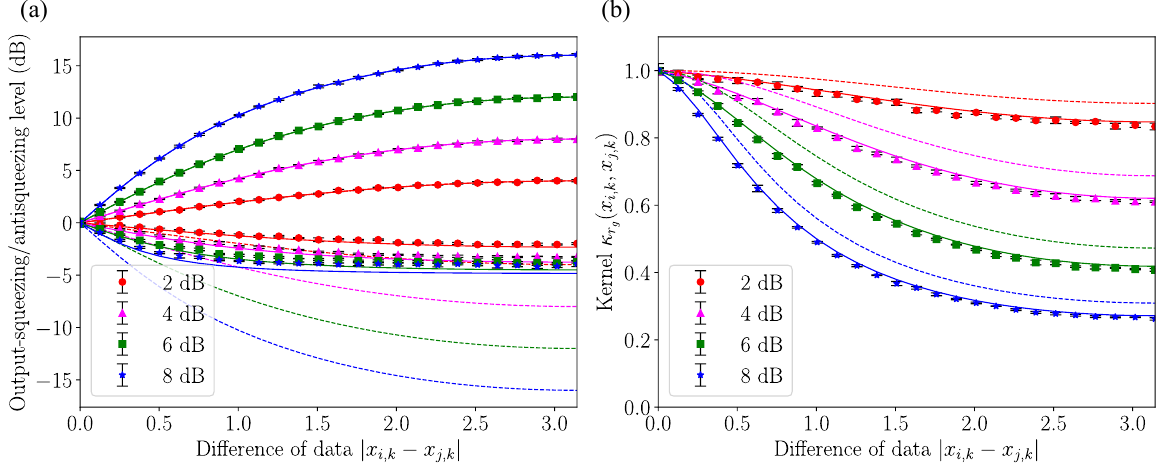}
\caption{The experimental result of the output-state measurement. We measure each point five times and plot the average and standard deviation. (a) The experimental result of the output-squeezing and antisqueezing levels (markers). We also show the theoretical kernel with (solid line) and without (dotted line) imperfections of our experiment. The red, magenta, green, and blue lines are the result of gate-squeezing levels  2, 4, 6, and 8 dB, respectively. (b) The experimental result of the squeezing-phase kernel (markers).}\label{fig:KernelAll}
\end{figure*}

\section{Experimental setup and analysis} Here we describe the details of our experimental setup in Fig.~\ref{fig:setup}(d), which is based on our previous experiment~\cite{ths:OurQAOA}. We use a continuous-wave laser of wavelength $1545.3\,\mathrm{nm}$.
The ancillary squeezed state is made with an optical parametric oscillator (OPO). The OPO is pumped by the second harmonic fields with wavelength of $772.7\,\mathrm{nm}$.
The pump power is set to $200\ \mathrm{mW}$.
The full width at half maximum of the OPO is $60\ \mathrm{MHz}$.\par

The variable beam splitter is composed of a bulk electro-optic modulator named EOM-1, a quarter-wave plate, and a pair of polarizing beam splitters.
EOM-1 serves as a variable polarization rotator and thus works as a variable beam splitter with the polarization optics.
We insert the quarter-wave plate so that the transmissivity is $50\,\%$ when no voltage is applied to EOM-1, which makes it easy to lock the relative phase between the squeezed light and the local oscillator fields at the homodyne detectors.

Each homodyne detection is performed by interfering the signal field with the local oscillator field at a 50:50 beam splitter.
Two beams from the beam splitter are received by two photodiodes. Then, the photocurrents are subtracted from each other and amplified in the electric circuit.
The bandwidth of the circuit is about $200\ \mathrm{MHz}$.
The optical power of the local oscillator field is set to $5\ \mathrm{mW}$.
A fiber-coupled electro-optic modulator EOM-2 shifts the optical phase of the local oscillator at HD2 for the control of the homodyne angle $\phi$. We measure the output state for $\phi=0,\ \pi/4,\ \pi/2$ to estimate the average and covariance matrix of the quadrature amplitudes, as is described in Ref.~\cite{ths:CovMat}. Because the output state is Gaussian, we can obtain its full information utilizing the average and covariance matrix.

The outcome of the homodyne detection is acquired by an oscilloscope and then sent to the classical computer.
The time-series waveform from the oscilloscope is converted to a quadrature amplitude by multiplying it by a mode function $h(t)$ defined by
\begin{equation}
h(t) = 
\begin{cases}
t\,\mathrm{e}^{-\Gamma^2t^2}&(|t|<t_1)\\
0&(\text{otherwise})
\end{cases},
\end{equation}
where $\Gamma=3\times10^7\,\mathrm{/s}$ and $t_1=50\,\mathrm{ns}$.
The purpose of using this mode function is to eliminate undesirable effects of low-frequency electrical noise from homodyne detectors \cite{ths:modefunc}.\par

Next, we explain the procedure to evaluate squeezing-phase kernel $\kappa_{r_g}(x_{i,k},\ x_{j,k})$ based on the above experimental setup. We first set the voltages applied to EOM-1, which corresponds to setting transmissivity $T$ of the variable beam splitter, according to input data $(x_{i,k},\ x_{j,k})$ and gate-squeezing level $r_g$. After setting transmissivity $T$, we measure the output state with homodyne angle $\phi$ set to $0,\ \pi/4,\ \pi/2$. From these results, we retrieve the covariance matrix and calculate the vacuum component of output state $\hat{S}(r_\mathrm{total}^{(\bm{i})})\ket{0}$. The calculated vacuum component corresponds to squeezing-phase kernel $\kappa_{r_g}(x_{i,k},x_{j,k})$. We acquire the squeezing-phase kernel by implementing the procedure above while changing transmissivity $T$ according to input data $(x_{i,k},x_{j,k})$.

As a preliminary measurement, the squeezed state from the OPO is measured by HD1 (HD2) with the variable beam splitter set to minimally (maximally) reflecting. The ancilla-squeezing and antisqueezing levels are measured to be $-4.8\,\mathrm{dB}$ ($-5.0\,\mathrm{dB}$) and $+8.9\,\mathrm{dB}$ ($+8.9\,\mathrm{dB}$) at HD1 (HD2).
This measurement result indicates that the overall optical loss of the experimental setup is $25\,\%$ ($23\,\%$) at HD1 (HD2) and the pure squeezing level is 10 dB. We use these parameters for the later analysis.

\begin{figure*}[t]
\centering
\includegraphics[width=1\linewidth]{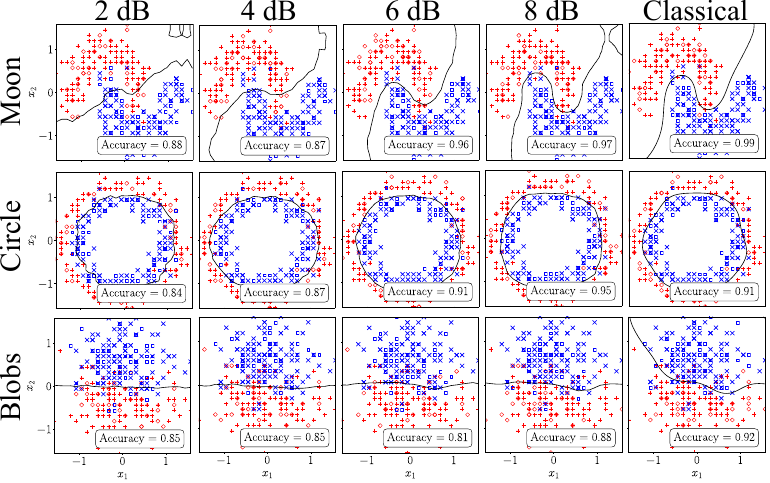}
\caption{The typical border with the experimental kernel. We show three types of datasets, moon (upper row), circle (middle row), and blobs (bottom row). The red and blue markers represent input data $(x_{i,1}, x_{i,2})$ with label $y_i=+1$ and $-1$, respectively. In each panel, plus and cross markers represent the training data, and square and diamond markers represent the test data. The black line indicates the classification border, which determines the prediction of the label with the unseen new inputs. Each of these datasets is classified by using the squeezing-phase kernel measured in our experiment with the gate-squeezing levels of 2, 4, 6, and 8 dB. The classification results with the classical kernel (Gaussian RBF kernel) are also shown in the rightmost column as a reference. The labels of the horizontal (vertical) axes of each panel are the same as those of the bottom (leftmost) panels. The regularization parameter $C$~\cite{ths:libsvm} is set to be 1.0 (default setting) for the support vector machine with the quantum and classical kernels. The hyperparameter $\gamma$~\cite{ths:libsvm} of the Gaussian RBF kernel is set to be 3.0, where the accuracy is almost saturated at the maximum for all types of the datasets.}\label{fig:border}
\end{figure*}

\begin{figure*}[t]
\centering
\includegraphics[width=1\linewidth]{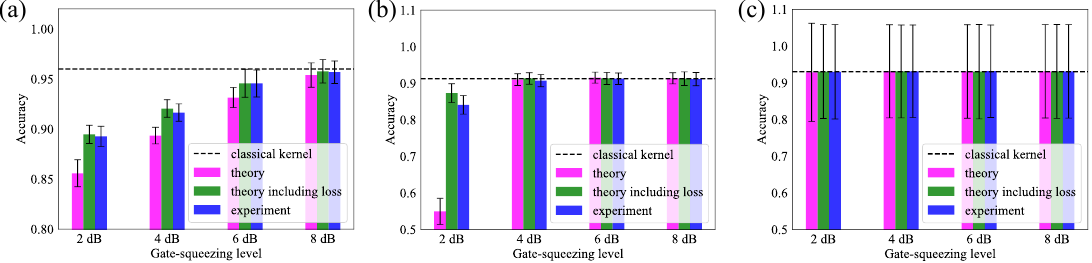}
\caption{The machine-learning results of accuracy with squeezing-phase kernel. The accuracy calculated from K-fold validation is shown for four different gate-squeezing levels. The error bar represents the standard deviation of learning with ten datasets. In each dataset, we calculated accuracies from three types of the kernel; theoretical kernel including experimental imperfections such as finite squeezing of the ancillary squeezed state and optical loss (theory including loss), theoretical kernel without including any experimental imperfections (theory), and experimental kernel (experiment). The black dotted line represents the accuracy obtained from the classical Gaussian-RBF kernel (classical kernel). The experimental result agrees well with the theoretical result including imperfections. Figures (a), (b), and (c) represent the accuracy of the moon, circle, and blobs datasets, respectively. The settings of the parameters $C$ and $\gamma$ are the same as in Fig.~\ref{fig:border}.}\label{fig:accuracy}
\end{figure*}

\section{Experimental result} The classification task is performed as follows: first, we prepare and pre-process $300$ input data, which is sufficiently large so as not to decrease accuracy of classification (details of pre-processing are described in the fourth paragraph in this section); second, we divide the input data into $225$ training and $75$ test data; third, we calculate squeezing-phase kernel $K_{r_g}(\bm{x}_i,\bm{x}_j)$ with each pair of training data $(\bm{x}_i,\bm{x}_j)$ using our quantum processor; fourth, the classical computer generates a classification model using the kernel handed from the quantum processor; finally, we classify the test data using the classification model and kernel between the training and test data. To experimentally demonstrate the successful implementation of our quantum kernel method, we show three results: (i) the squeezing-phase kernel measured in our experiment, (ii) the typical classification border drawn from our squeezing-phase kernel, and (iii) the accuracy calculated using K-fold method ($K=4$)~\cite{ths:kfold}. Each result of (i), (ii), and (iii) is shown in the following.\par

First, to qualitatively examine that our quantum processor evaluates properly the squeezing-phase kernel in Eq.~\eqref{eq:KernelExp}, we evaluate the squeezing and antisqueezing levels of the output state, what we call \textit{output-squeezing and antisqueezing levels}, for $r_g$ corresponding to 2, 4, 6 and 8 dB. Figure~\ref{fig:KernelAll}(a) shows the dependency of the output-squeezing level on the difference of data $|x_{i,k}-x_{j,k}|$. The output-squeezing and antisqueezing levels are measured at 26 points from zero to $\pi$. This range is sufficient to evaluate the kernel because the kernel is an even periodic function with a period of $2\pi$.  Each datum is evaluated from $3\times10000$ samples of $\hat{q}_{\mathrm{out},\phi}$ with $\phi=0,\ \pi/4,\ \pi/2$. The experimental results shown in Fig.~\ref{fig:KernelAll}(a) agree well with the theoretical ones which include experimental imperfections such as the finite ancilla-squeezing level and optical loss. The fact that the output-squeezing levels reach below zero means that our experiment is implemented in a quantum region. The saturation of the output-squeezing level around $5$ dB is due to $\sim5$ dB squeezing of ancillary squeezed states.\par

Then, we evaluate squeezing-phase kernel $\kappa_{r_g}(x_{i,k},x_{j,k})$ by calculating the vacuum component of the output state from these results, as shown in Fig.~\ref{fig:KernelAll}(b). The kernel structure of the experiment and the theory including the incompleteness of the experimental system agree well with each other. This means that experimental imperfections other than the ones included in the theoretical line, such as the limited accuracy of changing the transmissivity $T$, are sufficiently small so as not to influence the shape of the kernel in this experiment.\par

Next, we implement the classification task with a certain dataset to visualize how well our quantum kernel classifies the data, employing the support-vector-machine algorithm in scikit-learn~\cite{ths:libsvm} for solving Eq.~\eqref{eq:dualprob}. We use three types of datasets, called ``blobs,'' ``circle,'' and ``moon,'' generated by functions available in the scikit learn package. A dataset consists of pairs of input data $\bm{x}_i=(x_{i,1},x_{i,2})$ and the label of the input data $y_i\in\{-1,1\}$. We then pre-process the input data. First, $\bm{x}_i$ is standardized in $[-\pi/2, \pi/2]^2$ so that all the difference of data satisfies $0\leq|x_{i,k}-x_{j,k}|\leq\pi$. For experimental simplicity, we also discretize the generated $(x_{i,1},x_{i,2})$ to be on $26\times26$ lattice points. This pre-processing enables us to implement the classifications using only $26$ points of the pre-obtained kernel, which is shown in Fig.~\ref{fig:KernelAll}(b). The lattice is fine enough to retain accuracy compared to the accuracy without lattice-like pre-processing. We note this strategy is also used in the previous work~\cite{ths:DV_QML_handai}. The results are shown in Fig.~\ref{fig:border}. Figure~\ref{fig:border} indicates that the squeezing-phase kernel with the higher gate-squeezing level draws the decision boundary better so that the squeezing-phase kernel captures the feature of each dataset, comparably to the classical kernel.\par

Finally, to evaluate the performance of our quantum kernel quantitatively, we calculate the accuracy of classification from 300 input data using the K-fold method ($K=4$). We first divide the dataset into four subsets, each of which contains $75$ data. Then, we train the classifier using the $75\times3$ data and test its performance using the rest of the data ($75\times1$ data). Changing the subset for testing, we repeat the above procedure four times and obtain four accuracies. We also change the way of division ten times and obtain $4\times10$ accuracies per one dataset. We implement this calculation for ten different datasets and derive the average and the standard deviation of them. The results are shown in Fig.~\ref{fig:accuracy}. Figure~\ref{fig:accuracy} indicates that the classification of blobs datasets is well performed by the squeezing-phase kernel for gate-squeezing levels of 2, 4, 6, and 8 dB. On the other hand, the classifications of the moon and circle datasets are poor for lower gate squeezing levels, while they work well for higher gate squeezing levels. We attribute this behavior to the more nonlinear boundary of the moon and circle datasets compared with the blobs dataset, which requires a more nonlinear kernel and thus a higher gate squeezing level. We also see that the experimental accuracies agree well with theoretical ones including experimental incompleteness. Furthermore, the accuracies obtained from the squeezing-phase kernels for higher gate-squeezing levels are comparable to those obtained from the classical Gaussian-RBF kernel (black dotted line). From these results, we conclude that our implementation works as expected and is comparable to the classification with the classical kernel.\par

\section{Discussion} In conclusion, we demonstrated the successful implementation of the classification task with the CV quantum kernel method using the programmable photonic quantum processor. The results experimentally showed that the performances of the quantum kernel method were comparable to the classical one and agreed well with the numerical simulation. During the learning process, we mapped the data to the squeezed state. With this mapping, we demonstrated that we could implement the quantum kernel method with a single-mode CV quantum processor.\par

In this demonstration, even though the experimental system was influenced by various imperfections including optical loss, the quantum kernel method still successfully classified the data by tuning the gate-squeezing level. Such imperfections limit the effective ancilla-squeezing levels, affecting the shape of the quantum kernel. In addition, the gate-squeezing level also affects the shape. The optimum effective ancilla- and gate-squeezing levels for the overall performance depend on the given datasets and are nontrivial. In fact, with lower ancilla-squeezing levels or with higher gate-squeezing levels, the shape of the kernel in Fig.~\ref{fig:KernelAll}(b) becomes more nonlinear, indicating that the feature map becomes more nonlinear, and the kernel becomes more suitable for nonlinear separation. On the other hand, it is also observed in our numerical simulation that the accuracies of the quantum kernel method decrease with an extremely high gate-squeezing level of above 30 dB. This is probably because the peak of the kernel gets too sharp and the kernel function becomes flat at almost all region other than $|x_{i,k}-x_{j,k}|=0$ with such high squeezing, and the kernel cannot capture the feature of data. The optimum ancilla- and gate-squeezing levels should be investigated, but their detailed analysis is left for future work. The limited accuracy of changing the transmissivity would also change the shape of the kernel because it causes the difference of $r_{\mathrm{total}}^{(\bm{i})}$ and the insufficient cancellation of anti-squeezed component $\hat{p}_{\mathrm{a}}$ in the measurement-induced squeezing gate. This effect is sufficiently suppressed in this experiment but can become significant depending on experimental parameters.\par

In our demonstration, we use the two-dimensional datum $\bm{x}$. However, our method can be straightforwardly adapted to higher-dimensional data by increasing the range of $k$ in Eq.~\eqref{eq:kernel}. This increases the number of measurements, but such an increase can be avoided by discretizing the data and using the pre-obtained kernel repeatedly, as we did in our experiment. This strategy works because we encode each component of the data into a single-mode quantum state separately. In contrast, we can also consider the mapping of data to multi-mode entangled states~\cite{ths:EntangledKernel}, but it is left for future work.

Though our experiment in this paper can be simulated efficiently with classical computers because our system is entirely built with Gaussian building blocks~\cite{ths:GaussianSim2}, our implementation can be extended to a non-Gaussian regime by using non-Gaussian quantum states as feature maps. For example, a theoretical work~\cite{ths:nonClassKernel2} investigated the use of Fock states as feature maps. Another paper~\cite{ths:CV_QKM_nonClass} also investigated the nonclassicality witness of the CV quantum kernel by taking a quantum kernel based on single photon states as an example. As a next step, introducing such non-Gaussianity would be important to achieve a quantum advantage in the CV quantum kernel method~\cite{ths:QML_adv_1, ths:QML_adv2, ths:GeneralAdvQML}.

This demonstration sheds light on the strength of CV quantum computing. Our paper also stimulates the realizations of other QMLs in CV systems~\cite{ths:CV_Neural, ths:reservoir, ths:CV_linear} and thus opens a promising way toward quantum advantage. 

\section{Acknowledgments} This work was partly supported by
JST Grant Numbers JPMJFR223R and JPMJPF2221,
JSPS KAKENHI Grant Numbers 23H01102 and 23K17300,
the Canon Foundation,
and MEXT Leading Initiative for Excellent Young Researchers.
K. A. and S. I. acknowledge support from FoPM, WINGS Program, the University of Tokyo.
The authors thank Takahiro Mitani for the careful proofreading of the manuscript.

\appendix

\section{BLOCH MESSIAH DECOMPOSITION OF OUR CIRCUIT}\label{apa}

Here, we derive Eq.~\eqref{eq:r_total}. With the equations $\hat{S}(r,\theta)=\hat{R}(-\theta/2)\hat{S}(r,0)\hat{R}(\theta/2)$ and $\hat{R}(\phi)\ket{0}=\ket{0}$, $\kappa_{r_g}(x_{i,k},x_{j,k})$ can be described as 
\begin{align}
    \kappa_{r_g}(x_{i,k},x_{j,k})=&\|\braket{0|\hat{S}^\dagger(r_g,x_{i,k})\hat{S}(r_g,x_{j,k})|0}\|^2\nonumber\\
=&\|\bra{0}\hat{R}\left(-\frac{x_{i,k}}{2}\right)\hat{S}(-r_g,0)\hat{R}\left(\frac{x_{i,k}}{2}\right)\nonumber\\
&\hat{R}\left(-\frac{x_{j,k}}{2}\right)\hat{S}(r_g,0)\hat{R}\left(\frac{x_{j,k}}{2}\right)\ket{0}\|^2\nonumber\\
=&\|\braket{0|\hat{S}(-r_g,0)\hat{R}(\Delta^{(\bm{i})})\hat{S}(r_g,0)|0}\|^2,\\
\Delta^{(\bm{i})}=&\frac{x_{i,k}-x_{j,k}}{2}.
\end{align}
Then, the transformation matrix $M$ of the quadrature amplitudes with the unitary operator $\hat{S}(-r_g,0)\hat{R}(\Delta^{(\bm{i})})\hat{S}(r_g,0)$ can be described as
\begin{align}
    M&=\begin{bmatrix}
    e^{r_g} & 0\\
    0 & e^{-r_g}\\
    \end{bmatrix}
    \begin{bmatrix}
    \cos\Delta^{(\bm{i})} & \sin\Delta^{(\bm{i})}\\
    -\sin\Delta^{(\bm{i})} & \cos\Delta^{(\bm{i})}\\
    \end{bmatrix}
    \begin{bmatrix}
    e^{-r_g} & 0\\
    0 & e^{r_g}\\
    \end{bmatrix}\nonumber\\
&=
    \begin{bmatrix}
    \cos\Delta^{(\bm{i})} & e^{2r_g}\sin\Delta^{(\bm{i})}\\
    -e^{-2r_g}\sin\Delta^{(\bm{i})} & \cos\Delta^{(\bm{i})}\\
    \end{bmatrix}.
\end{align}
We apply Bloch-Messiah decomposition \cite{ths:bloch} to the matrix $M$. The eigenvalues of $MM^\dagger$ and $M^\dagger M$ are both
\begin{align}
    \lambda_{\pm}^{(\bm{i})}&=\cos^2\Delta^{(\bm{i})}+\cosh(4r_g)\sin^2\Delta^{(\bm{i})}\nonumber\\
&\pm\left[\sinh^2(4r_g)\sin^4\Delta^{(\bm{i})}\right.\nonumber\\
&\left.+4\sinh^2(2r_g)\sin^2\Delta^{(\bm{i})}\cos^2\Delta^{(\bm{i})}\right]^{\frac{1}{2}}.
\end{align}
With this eigenvalue, the matrix $M$ is decomposed into
\begin{align}
    M&=
    \begin{bmatrix}
    \cos\phi_1^{(\bm{i})} & \sin\phi_1^{(\bm{i})}\\
    -\sin\phi_1^{(\bm{i})} & \cos\phi_1^{(\bm{i})}\\
    \end{bmatrix}\nonumber\\
    &\begin{bmatrix}
    \frac{1}{\sqrt{\lambda_+^{(\bm{i})}}}& 0\\
    0 & \sqrt{\lambda_+^{(\bm{i})}}\\
    \end{bmatrix}
    \begin{bmatrix}
    \cos\phi_2^{(\bm{i})} & \sin\phi_2^{(\bm{i})}\\
    -\sin\phi_2^{(\bm{i})} & \cos\phi_2^{(\bm{i})}\\
    \end{bmatrix},\label{eq:M}
\end{align}
which corresponds to the operator $\hat{R}(\phi_1^{(\bm{i})})\hat{S}(\mathrm{ln}\lambda_+^{(\bm{i})}/2,0)\hat{R}(\phi_2^{(\bm{i})})$. Therefore, we can define the squeezing parameter by $r_\mathrm{total}^{(\bm{i})}=\mathrm{ln}\lambda_+^{(\bm{i})}/2$, which is equivalent to Eq.~\eqref{eq:r_total}.\\

\section{INPUT-OUTPUT RELATION OF THE OPTICAL CIRCUIT}\label{apb}

Here, we derive Eq.~\eqref{eq:output}. To perform the squeezing operation $\hat{S}(r_\mathrm{total}^{(\bm{i})},0)$, we use a measurement-induced squeezing gate~\cite{ths:univsq} shown in Fig.~\ref{fig:setup}(c). First, we explain how this measurement-induced implementation in Fig.~\ref{fig:setup}(c) works. The quadrature amplitudes of an input vacuum state and an ancillary $x$-squeezed state are described as $(\hat{q}_\mathrm{in},\hat{p}_\mathrm{in})$ and $(\hat{q}_\mathrm{a},\hat{p}_\mathrm{a})$, respectively. Let $(\hat{q}_i,\hat{p}_i)$ and $(\hat{q}_{i,\phi},\hat{p}_{i,\phi})$ ($i=1,2,\mathrm{out}$, and $\phi\in \mathbb{R}$) denote the quadrature amplitudes of the beams coming out of the beam splitter defined by
\begin{align}
\hat{q}_1&=\sqrt{1-T}\hat{q}_\mathrm{in}+\sqrt{T}\hat{q}_\mathrm{a},\\
\hat{p}_1&=\sqrt{1-T}\hat{p}_\mathrm{in}+\sqrt{T}\hat{p}_\mathrm{a},\\
\hat{q}_2&=\sqrt{T}\hat{q}_\mathrm{in}-\sqrt{1-T}\hat{q}_\mathrm{a},\\
\hat{p}_2&=\sqrt{T}\hat{p}_\mathrm{in}-\sqrt{1-T}\hat{p}_\mathrm{a},\\
\begin{bmatrix}
\hat{q}_{i,\phi}\\
\hat{p}_{i,\phi}
\end{bmatrix}
&=
\begin{bmatrix}
\cos\phi&\sin\phi\\
-\sin\phi&\cos\phi\
\end{bmatrix}
\begin{bmatrix}
\hat{q}_i\\
\hat{p}_i
\end{bmatrix}.
\end{align}
After the beam splitter operation, we measure $\hat{p}_1$ using the homodyne detection and multiply the result by the feedforward gain $g$, and add it to $\hat{p}_2$. By choosing $g=\sqrt{(1-T)/T}$ to counteract the antisqueezed component $\hat{p}_a$, the output state becomes 
\begin{equation}\label{eq:tochu}
\begin{bmatrix}
\hat{q}_\mathrm{out}\\
\hat{p}_\mathrm{out}\\
\end{bmatrix}
=
\begin{bmatrix}
\hat{q}_{2}\\
\hat{p}_{2}+g\hat{p}_1\\
\end{bmatrix}\\
=
\begin{bmatrix}
\sqrt{T}\hat{q}_\mathrm{in}\\
\frac{1}{\sqrt{T}}\hat{p}_\mathrm{in}\\
\end{bmatrix}\\
-
\begin{bmatrix}
\sqrt{1-T}\hat{q}_a\\
0\\
\end{bmatrix}\\.
\end{equation}
In the high squeezing limit of $\hat{q}_a\rightarrow0$, the output state becomes
\begin{equation}
\begin{bmatrix}
\hat{q}_\mathrm{out}\\
\hat{p}_\mathrm{out}\\
\end{bmatrix}
=
\begin{bmatrix}
\sqrt{T}\hat{q}_\mathrm{in}\\
\frac{1}{\sqrt{T}}\hat{p}_\mathrm{in}\\
\end{bmatrix}\\,
\end{equation}
which means a squeezing operation is performed on the input state.
Here, to set the squeezing level as $r_\mathrm{total}^{(\bm{i})}$, we define $T = \exp(-2r_\mathrm{total}^{(\bm{i})})$. Then $\hat{q}_{\mathrm{out},\phi}$ is calculated as
\begin{align}
    \hat{q}_{\mathrm{out},\phi}
    &=\hat{q}_\mathrm{out}\cos\phi + \hat{p}_\mathrm{out}\sin\phi\nonumber\\
    &=e^{-r_\mathrm{total}^{(\bm{i})}}\hat{q}_\mathrm{in}\cos\phi+e^{r_\mathrm{total}^{(\bm{i})}}\hat{p}_\mathrm{in}\sin\phi\nonumber\\
    &-\sqrt{1-\exp(-2r_\mathrm{total}^{(\bm{i})})}\hat{q}_\mathrm{a}\cos\phi,\label{eq:XoutPhi}
\end{align}
which coincides with Eq.~\eqref{eq:output}.\par

Then, we also note that the feedforward operation can be implemented by post-processing as Fig.~\ref{fig:setup}(d) shows. This can be explained by transforming $\hat{q}_{\mathrm{out},\phi}$ in Eq.~\eqref{eq:XoutPhi} as
\begin{align}
    \hat{q}_{\mathrm{out},\phi}
    &=\hat{q}_{2}\cos\phi + (\hat{p}_{2}+g\hat{p}_1)\sin\phi\nonumber\\
    &=(\hat{q}_{2}\cos\phi + \hat{p}_{2}\sin\phi) + g\sin\phi\cdot\hat{p}_1\nonumber\\
    &=\hat{q}_{2,\phi} + g\sin\phi\cdot\hat{p}_1.\label{eq:postprocess}
\end{align}
Hence, $\hat{q}_{\mathrm{out},\phi}$ can be obtained by first measuring $\hat{q}_{2,\phi}$ and $\hat{p}_1$, and then summing them up as Eq.~\eqref{eq:postprocess} shows. We adopt this strategy in Fig.~\ref{fig:setup}(d).

%

\end{document}